\documentclass[preprint,1p,times,authoryear]{elsarticle}

\usepackage{amssymb}

\usepackage{amsmath}
\usepackage{xcolor}
\usepackage[normalem]{ulem}
\usepackage{soul}
\usepackage[utf8]{inputenc}
\usepackage[T1]{fontenc}
\usepackage{listings}

\journal{Expert Systems With Applications}

\begin{document}

\begin{frontmatter}



\title{Multi-objective Portfolio Optimization Via Gradient Descent} 


\author[label1]{Christian Oliva}
\author[label2]{Pedro R. Ventura}
\author[label1]{Luis F. Lago-Fernández}

\affiliation[label1]{organization={Grupo de Neurocomputación Biológica, Departamento de Ingeniería Informática, Escuela Politécnica Superior, Universidad Autónoma de Madrid},
            addressline={c/ Francisco Tomás y Valiente, 11}, 
            city={Madrid},
            postcode={28049}, 
            state={Madrid},
            country={Spain}}

\affiliation[label2]{organization={March Asset Management, S.G.I.I.C., S.A.U.},
                     addressline={Castello, 74},
                     city={Madrid},
                     postcode={28006},
                     state={Madrid},
                     country={Spain}}

\begin{abstract}
Traditional approaches to portfolio optimization, often rooted in Modern Portfolio Theory and solved via quadratic programming or evolutionary algorithms, struggle with scalability or flexibility, especially in scenarios involving complex constraints, large datasets and/or multiple conflicting objectives. To address these challenges, we introduce a benchmark framework for multi-objective portfolio optimization (MPO) using gradient descent with automatic differentiation. Our method supports any optimization objective, such as minimizing risk measures (e.g., CVaR) or maximizing Sharpe ratio, along with realistic constraints, such as tracking error limits, UCITS regulations, or asset group restrictions. We have evaluated our framework across six experimental scenarios, from single-objective setups to complex multi-objective cases, and have compared its performance against standard solvers like CVXPY and SKFOLIO. Our results show that our method achieves competitive performance while offering enhanced flexibility for modeling multiple objectives and constraints. We aim to provide a practical and extensible tool for researchers and practitioners exploring advanced portfolio optimization problems in real-world conditions.
\end{abstract}


\begin{highlights}
\item Gradient descent framework for multi-objective portfolio optimization
\item Supports any objective, such as risk minimization or Sharpe ratio maximization
\item Constraints are handled through regularization-inspired terms
\item Easily extensible to more complex MPO problems just by adding new constraints
\item Fully reproducible and accessible for researchers and practitioners
\end{highlights}


\end{frontmatter}



\section{Introduction}
\label{sec:introduction}

Portfolio optimization (PO) is a crucial aspect of financial management and investment planning with the aim of developing the best combination of having less risk and obtaining more profit in an investment. The most common investment strategy is to build a portfolio considering different securities to diversify risk. However, traditional portfolio analysis requires evaluating the return and risk conditions of individual securities, which may not be successful. This concept is rooted in the Modern Portfolio Theory and the Efficient Frontier, introduced by \cite{MarkowitzPortfolioSelectionArticle}. According to this theory, an investor attempts to maximize its portfolio's return for a given amount of risk, or vice-versa, minimize its risk for a given level of expected return. However, the Markowitz model has some drawbacks. It relies on historical price series and the covariance matrix, making it sensitive to input data and computationally challenging when dealing with a large number of assets. Over the years, this concept evolved into the Capital Asset Pricing Model (CAPM) of \cite{SharpeCapitalAssetPricesPaper} and \cite{LintnerValuationRiskAssetsArticle}, who introduced the Market Portfolio and the Sharpe ratio. Subsequently, new models have been introduced, such as the Black-Litterman model of \cite{BlackLitterman}, the Three Factors Model of \cite{FamaAndFrench3Factors}, and the Risk Parity and Budgeting of \cite{roncalli2014introductionriskparitybudgeting}, including research that allows us to stabilize the optimized portfolio, such as denoising the covariance matrix or shrinkage methods \citep{Introduction_to_Risk_Parity_Budgeting}.

On the other hand, recent research in portfolio optimization has proposed new risk measures such as the value-at-risk (VaR, \cite{VaR}), the Conditional Value-at-risk (CVar, \cite{CVaR}) or the Conditional Drawdown-at-risk (CDaR, \cite{CDaR}). These advancements have allowed for a more nuanced understanding of risk, catering to various investor preferences and constraints. Nowadays, portfolio optimization has evolved beyond this traditional single-objective framework of balancing risk and return. Investors and portfolio managers aim to address multiple and conflicting objectives simultaneously. For example, one might seek to minimize CVaR while maximizing the Sharpe ratio and imposing constraints such as limiting the number of assets in the portfolio or enforcing sector diversification. 

This approach reflects the complexity of real-world investment scenarios and has led to a multi-objective optimization (MO) scenario. MO is the problem of simultaneously optimizing two or more conflicting objectives with certain constraints. Thus, multi-objective portfolio optimization (MPO), which integrates the principles of MO into portfolio management, provides a robust framework for developing modern portfolio construction. However, there is no optimal solution that maximizes or minimizes each objective to its fullest in MPO \citep{Gandibleux2005198420042} since the various objective functions in the problem are usually in conflict with each other. Therefore, MPO aims to find efficient solutions that provide a trade-off between the different objectives \citep{zitzler_2000,zitzler_2002}.

During the last decades, MPO has attracted attention from academics and investors to address the growing complexity of financial markets. Researchers have developed advanced mathematical models and computational algorithms, such as evolutionary algorithms (EA) and machine learning (ML) techniques. Instead of using these complex models, this paper aims to present a benchmark for MPO based on gradient descent optimization. This approach performs efficient and scalable MPO, enabling the identification of optimal solutions by iteratively adjusting portfolio weights with the gradient descent technique. The automatic differentiation provided by Tensorflow \citep{tensorflow} and PyTorch \citep{pytorch}, two of the most popular Deep Learning frameworks, offer computational simplicity and adaptability to solve large-scale portfolio problems for real-world scenarios.

Our approach supports any financial objective and constraint, including regulatory rules (e.g., UCITS), risk metrics (e.g., CVaR), and practical portfolio construction rules (e.g., maximum tracking error allowed, asset limits, etc.). We show that this framework achieves competitive performance compared to standard optimization tools while offering greater modeling flexibility. In addition, we provide open-source implementations in \ref{ap:tf_differentiation_techniques} and \ref{ap:tf_constraints} to facilitate reproducibility\footnote{We provide a Github repository with the code used in this paper: https://github.com/pventura1976/mpo-gradient-descent} and future research.

The paper is organized as follows. Section \ref{sec:related_work} reviews the existing literature on portfolio optimization methods, with a focus on evolutionary and gradient-based techniques. Section \ref{sec:materials} describes our proposed framework in detail, including the optimizer, the formulation of the loss function, and the technical considerations for handling constraints in a differentiable setting to ensure compatibility with gradient descent. Section \ref{sec:results} presents experimental results across six scenarios, from simple objectives to complex multi-objective constrained problems. Finally, in section \ref{sec:conclusion}, we summarize the main conclusions and outlines directions for future work.

\section{Related Work}
\label{sec:related_work}

Exact solution algorithms can tackle PO problems since they have a quadratic structure. However, when considering additional constraints and objectives, or when the dimension of the problem increases, exact solution algorithms may face trouble while handling MPO problems \citep{kalayci_2019}. Recent research has used inexact techniques to solve more realistic MPO problems that incorporate non-convex constraints in the mathematical formulation, thus turning the problem NP-hard \citep{moral_2006}.

Traditional research is based on EAs, usually Genetic algorithms (GAs). These algorithms are based on emulating the mechanisms of natural selection to solve optimization problems. These are population-based stochastic optimization heuristics inspired by Darwin's evolution theory. GAs search through a solution space by evaluating possible solutions (individuals). They start with a random initial population, and then the fitness of individuals is determined by evaluating the objective function. The best individuals survive, and new individuals for the next generation are created by combining their parents and altering their genes through random mutations. This cycle repeats until a breaking criterion is completed. \cite{Schaffer_85,Schaffer_85_thesis,Schaffer_85_IJCAI} introduced the first implementation of a multi-objective evolutionary algorithm: the Vector Evaluation Genetic Algorithm (VEGA). In reality, it is just a simple genetic algorithm with a modified survivor selection mechanism, but they opened the way to MO with GAs. \cite{Fonseca1993,Horn1994ANP,Srinivas_94,Zitzler_99,Knowles_2000} followed this approach and proposed different multi-objective genetic algorithms that extend the traditional VEGA by using the Pareto domination ranking and fitness.

The ability of GAs to deal with a set of possible solutions makes them naturally suitable for MPO problems. \cite{Arnone_93} were the first to use GAs to optimize investment portfolios. They implemented the Markowitz model, but proposed to use lower partial moments as a risk measure. The use of downside risk makes the problem non-convex, so quadratic solvers can not find exact solutions. This opened a new line of research, and many works appeared a few years later. \cite{FosterJamesA1996AGas,Vedarajan_1997,Chang2000HeuristicsFC}, among others, consider the Markowitz model with different variations that incorporate many constraints. We suggest \cite{METAXIOTIS201211685} for a more extended review of traditional GAs applied to MPO.

However, while genetic algorithms are the most preferred evolutionary algorithms in mean-variance portfolio optimization, swarm-based algorithms (SAs) have increased in popularity when facing an MPO problem. SAs are based on the study of computational systems inspired by the behaviors of animals living in their natural environment. Swarms, such as flocks of birds or colonies of ants, reflect their cooperation in computational systems. \cite{Chen_2006} were the first to apply particle swarm optimization to the MPO problem. \cite{GARCIA201210722,Mishra_2016} proposed robust multi-objective swarm approaches and compared their results with traditional genetic algorithms, reporting that swarm optimization outperforms the others, providing the best Pareto optimal solutions. \cite{kalayci_2019} report that particle swarm optimization and artificial bee colony algorithms are the most popular SAs for the MPO problem.

Nevertheless, it is surprising that Machine Learning (ML), which has reached a high popularity, has not been extensively explored in the context of MPO problems. \cite{FERNANDEZ20071177} applied Hopfield networks and compared their results with GAs and SAs, reporting higher performance in larger instances. \cite{YU200834} proposed a neural network based on mean–variance–skewness and report high performance on MPO problems. \cite{kalayci_2019} show that more than 80\% of the research related to MPO uses metaheuristics (including swarm optimization and genetic algorithms), while ML approaches only cover a 12\%, \cite{deng2024milliongeneralmultiobjectiveframework}, have introduced a general multi-objective framework using reinforcement learning reporting improved profitability and risk resistance, as well as better generalization ability across different financial market. In summary, it seems that population-based algorithms, such as SAs and GAs, are winning the battle and dominating the field of MPO research due to their flexibility and ease of implementation. 

\section{Materials and Methods}
\label{sec:materials}

In this section, we present the benchmark for multi-objective portfolio optimization problems based on the gradient descent technique. This framework is designed to address the challenges of balancing multiple conflicting objectives in portfolio optimization, such as maximizing Sharpe ratio while minimizing CVaR with some constraints (see section \ref{subsubsec:max_sharpe_min_cvar_te}). We aim to provide a systematic approach to solve MPO problems, while maintaining computational efficiency. The following sections detail the optimization techniques (sec. \ref{subsec:optimizer}), the formulation of the multi-objective loss function (sec. \ref{subsec:multiobjective_loss}), some technical considerations for differentiation (sec. \ref{subsec:technical_considerations}), the definitions of the tested objectives and constraints (sec. \ref{subsec:objectives_and_constraints}), and the experimental setup used to validate the gradient descent benchmark (section \ref{subsec:experiments}).

\subsection{Gradient Descent Optimizer}
\label{subsec:optimizer}

Portfolio optimization is the process of selecting the best possible mix of assets to achieve specific financial goals, such as maximizing returns while minimizing risk. The most widely known method of portfolio optimization is the mean-variance optimization \citep{MarkowitzPortfolioSelectionBook}, which maximizes the expected return for a given level of risk by considering the expected return of the assets and their covariances:

\begin{equation}
    max \quad (\mathbb{E}[\mathbf{R}] - \lambda Var(\mathbf{R})), \label{eq:markowitz}
\end{equation}

\noindent where $\mathbf{R}$ represents the portfolio returns and:

\begin{equation}
    \mathbb{E}[\mathbf{R}] = \sum_{i=1}^n \mathbf{w}_i \mathbb{E}[\mathbf{r}_i] = \mathbf{w}^T \mathbb{E}[\mathbf{r}],
\end{equation}

\begin{equation}
    Var(\mathbf{R}) = \sum_{i=1}^n \sum_{j=1}^n \mathbf{w}_i Cov(\mathbf{r}_i, \mathbf{r}_j) \mathbf{w}_j = \mathbf{w}^T \Sigma \mathbf{w},
\end{equation}

\noindent over the constraint $\sum_{i=1}^n \mathbf{w}_i = 1$. In the previous equations, $\mathbb{E}[\mathbf{R}]$ is the expected return of the portfolio, $\mathbb{E}[\mathbf{r}_i]$ is the expected return of the $i$th asset in the portfolio, $\mathbf{w}_i$ is the weight of the $i$th asset in the portfolio, $\Sigma$ is the covariance matrix, and $n$ is the number of assets in the portfolio. The investor's risk aversion is reflected in the parameter $\lambda$. The higher the value of $\lambda$, the higher the risk aversion. For simplicity, we will not allow short positions; therefore, asset weights must be non-negative: $\mathbf{w}_i \ge 0 \quad \forall i$. Therefore, the portfolio optimization problem consists of determining the weights $\mathbf{w}_i$ of the portfolio that maximize equation \ref{eq:markowitz} with the two mentioned constraints.

However, in practice, we do not typically work with just expected returns; instead, we work with time series of returns for all assets, such as daily returns for each asset. Therefore, if we already know the weights $\mathbf{w}$ (or we are in the process of optimizing them), we can treat the portfolio return as another asset and estimate it using the following equation:

\begin{equation}
    R_t = \sum_{i=1}^n \mathbf{w}_i \mathbf{r}_{i,t},
\end{equation}

\noindent where $R_t$ is the return of the portfolio at time $t$, $\mathbf{r}_{i,t}$ is the return of the $i$th asset at time $t$, and $\mathbf{w}_i$ is its corresponding weight in the portfolio. This way, we can calculate the expected return and the variance of the portfolio in a simpler way:

\begin{equation}
    \mathbb{E}[\mathbf{R}] = \overline{\mathbf{R}} = \frac{1}{T} \sum_{t=1}^T R_t,
\end{equation}

\begin{equation}
    Var(\mathbf{R}) = \sigma^2(\mathbf{R}) = \frac{1}{T} \sum_{t=1}^T(R_t - \overline{\mathbf{R}})^2
\end{equation}

Once we have the return series $\mathbf{r}_{i,t}$, we can proceed with the optimization process to build the benchmark. First, we define the variables to be optimized, the pre-weights $\mathbf{z}$ for each asset in the portfolio, which are randomly initialized. In addition, to ensure that the portfolio weights are positive and sum to 1, we apply the softmax function to the pre-weights:

\begin{equation}
    \mathbf{w}_i = softmax_i(\mathbf{z}) = \frac{exp(\mathbf{z}_i)}{\sum_{j=1}^n exp(\mathbf{z}_j)}
\end{equation}

The softmax function converts the pre-weights $\mathbf{z}$ into a valid probability distribution, ensuring that the sum of the portfolio weights $\mathbf{w}$ equals 1 and that they are non-negative. This step allows us to perform gradient descent optimization while maintaining the constraints. However, softmax has an intrinsic limitation: the resulting probability distribution always has full support. In other words, $softmax_i(\mathbf{z}) \neq 0$ for almost every $\mathbf{z}$ and $i$. This is a disadvantage when a sparse probability distribution is desired, such as in the MPO problem. In these cases, it is common to define a threshold below which small probability values are truncated to zero. Thus, we propose to apply an alternative to the softmax function: the sparsemax \citep{sparsemax_martins16}, which is defined as follows:

\begin{equation}
    \mathbf{w} = sparsemax(\mathbf{z}) = [\mathbf{z} - \tau(\mathbf{z})]_+,
\end{equation}

\noindent where $[\cdot]_+ = \max(0, \cdot)$, and $\tau(\mathbf{z})$ is the threshold function that satisfies $\sum_{j=1}^n \mathbf{w}_i = 1$ for every $\mathbf{z}$, which is expressed as:

\begin{equation}
    \tau(\mathbf{z}) = \frac{\left(\sum_{j\leq k(\mathbf{z})} \mathbf{z}_j\right)-1}{k(\mathbf{z})},
\end{equation}

\noindent where $k(\mathbf{z}) := \max \{ k \in [K] | 1 + k \mathbf{z}_k > \sum_{j\leq k} \mathbf{z}_j\}$. In other words, the sparsemax function defines a threshold $\tau_i(\mathbf{z})$ for each pre-weight $\mathbf{z}_i$, so that the components lower than its corresponding threshold will be truncated to zero, and the remaining weights sum to 1. In \ref{ap:softmax_vs_sparsemax} we compare the softmax and the sparsemax functions with the same weights vector. With this setup, we can now define the following loss function for the optimization:

\begin{equation}
    L(\mathbf{z}, \lambda) = - (\overline{\mathbf{R}} - \lambda Var(\mathbf{R})), \label{eq:loss_markowitz}
\end{equation}

\noindent which is equivalent to the proposal of \cite{MarkowitzPortfolioSelectionBook} (see equation \ref{eq:markowitz}). Note the minus sign in the loss function as we aim to maximize the objective. The optimization procedure applies gradient descent to optimize the portfolio pre-weights $\mathbf{z}$. It iteratively adjusts $\mathbf{z}$ to minimize the loss function $L$ (maximize the objective due to the minus sign), and the sparsemax (also works with softmax) function ensures that the portfolio weights remain valid throughout the iterations.

\subsection{Multi-objective Loss}
\label{subsec:multiobjective_loss}

Up to this point, we have presented a gradient descent optimizer for the Markowitz mean-variance model. This framework provides a foundational approach to portfolio optimization, but operates under a single-objective paradigm. However, in real-world scenarios, investors face multiple conflicting objectives, adding various constraints such as budget, sector exposure, or regulatory requirements. To address these complexities, MPO extends the traditional Markowitz framework employing weighted terms (similarly to the regularization penalty in the loss function for neural networks) to introduce and manage restrictions effectively. This allows us to incorporate these constraints directly into the loss function:

\begin{equation}
    L(\mathbf{z}, \lambda, \lambda_1, \lambda_2, ..., \lambda_N) =  - (\overline{\mathbf{R}} - \lambda \, \mathrm{Var}(\mathbf{R})) + \sum_{i=1}^N \lambda_i C_i
\end{equation}

\noindent where $C_i$ represents the $i$th objective or constraint, and $\lambda_i$ is the multiplier associated with $C_i$. These multipliers reflect the relative importance of each objective, allowing the model to explore the trade-off between these multiple objectives. Note that the objective function could be replaced by any other of interest. For all other aspects, the gradient descent optimizer described in the previous section remains the same, and the only thing that must be modified is the loss function according to our needs.

\subsection{Technical considerations for differentiation}
\label{subsec:technical_considerations}

When implementing the constraints, we have to pay attention to the technical nuances of differentiation. Certain constraints introduce non-differentiable operations that interfere with the gradient-based optimization framework. This section addresses two common situations that require special care: the enforcement of threshold constraints (sec. \ref{subsubsec:thresholds}), and the use of conditional masks for asset selection (sec. \ref{subsubsec:mask}). In both cases, we implement a technique that maintains compatibility with our optimizer. For reproducibility, the Tensorflow implementation code can be found in \ref{ap:tf_differentiation_techniques}.

\subsubsection{Differentiation with threshold constraints}
\label{subsubsec:thresholds}

Many constraints ensure that a variable remains below a certain threshold. For example, an asset weight $\mathbf{w}_i$ cannot exceed $10\%$ because we want to force some diversification ($w_i \leq 0.1 \quad \forall i$). This kind of constraints, which include a threshold $\alpha=0.1$, makes the processing not differentiable. To solve this, we introduce a penalty term $C$ that activates when the weights $\mathbf{w}$ exceed the threshold $\alpha$, as follows:

\begin{equation}
    C = [\mathbf{w}-\alpha]_+ = ReLU(\mathbf{w}-\alpha), \label{eq:threshold_constraint}
\end{equation}

\noindent where $[\cdot]_+ = \max(0, \cdot) = ReLU(\cdot)$. This constraint is non-zero when any $\mathbf{w}_i > \alpha$, so the loss function is penalized when this happens.

\subsubsection{Differentiation of asset conditional masks}
\label{subsubsec:mask}

\begin{figure}[!bht]
    \centering
    \includegraphics[width=\linewidth]{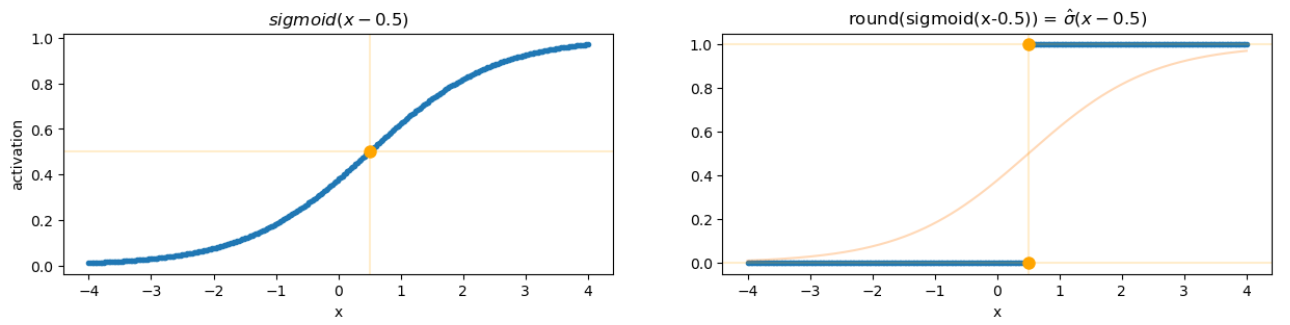}
    \caption{Round effect on the sigmoid activation function, following the equation \ref{eq:round_sigmoid}. Note that the vertical orange line is exactly at $\alpha = 0.5$, where the mask switches its value.}
    \label{fig:sigmoid_vs_round_sigmoid}
\end{figure}

In many MPO problems, conditional masks are often necessary, for instance, in asset selection. For example, if we want to select assets whose weights exceed $50\%$, we can introduce a conditional mask based on a sigmoid function:

\begin{equation}
    mask = \hat{\sigma}(x - \alpha) = round(sigmoid(x - \alpha)) \label{eq:round_sigmoid}
\end{equation}

\noindent where $x$ is the input variable and $\alpha$ is a threshold parameter ($0.5$ in the example). When $x > \alpha$, the rounded sigmoid function is equal to $1$, and otherwise it is $0$, as shown in figure \ref{fig:sigmoid_vs_round_sigmoid}. The $round$ function ensures that the mask takes binary values, effectively converting the smooth sigmoid into a boolean mask. 

This function, of course, presents a problem when applying gradient descent: the $round$ function is not differentiable. However, this issue has an easy solution if, for gradient computation during training, we use the derivative of the original sigmoid function before rounding:

\begin{equation}
    sigmoid(x) (1-sigmoid(x)) \label{eq:derivative_round_sigmoid}
\end{equation}

This mask (eq. \ref{eq:round_sigmoid}) allows the optimizer to filter assets under specific conditions while maintaining differentiability during optimization. Gradient updates remain well defined, enabling automatic differentiation through backpropagation, and ensuring that asset constraints are respected while preserving the flexibility of our gradient-based optimization framework. We show the Tensorflow implementation of the $round$ function and its derivative in Table \ref{code:round_sigmoid}, and the mask in Table \ref{code:mask_threshold}.

In summary, these two methods (equations \ref{eq:threshold_constraint} and \ref{eq:round_sigmoid}), together with the basic mathematical operations, are sufficient to handle the typical constraints in MPO problems without requiring more complex adjustments, as shown in the following section (sec. \ref{subsec:objectives_and_constraints}). The constraints are formulated to penalize the loss function, guiding the optimizer to learn how to satisfy the restrictions effectively.

\subsection{Objectives and constraints}
\label{subsec:objectives_and_constraints}

This section provides the primary objectives and constraints considered in the evaluation of our gradient descent optimizer benchmark for MPO. Following the technical considerations for differentiation described in section \ref{subsec:technical_considerations}, we can easily define different objectives, such as maximizing Sharpe ratio (sec. \ref{subsubsec:obj_sharpe}) or minimizing CVaR (sec. \ref{subsubsec:obj_cvar}), and also different constraints: tracking error limitations (sec. \ref{subsubsec:cons_tracking_error}), the adherence to a simplified version of the UCITS directive (sec. \ref{subsubsec:cons_ucits}), and conditions on asset weights (sec. \ref{subsubsec:cons_weights_min}) and their active range (sec. \ref{subsubsec:cons_range}). By combining and balancing these objectives and constraints, we aim to consolidate the efficacy of the optimizer for MPO problems. The Tensorflow implementation of these objectives and constraints can be found in \ref{ap:tf_constraints}.

\subsubsection{Objective: Sharpe ratio}
\label{subsubsec:obj_sharpe}

Instead of generating the entire set of efficient portfolios by varying the risk aversion parameter $\lambda$ in the standard mean-variance optimization formulation (see equation \ref{eq:markowitz}), we directly search for the efficient portfolio that maximizes the risk-return trade-off. This is achieved by maximizing the Sharpe ratio, a widely used metric in finance that evaluates portfolio performance relative to its risk. The Sharpe ratio quantifies the difference between the expected return of the portfolio and the risk-free rate, relative to the portfolio's risk, which is represented by its standard deviation:

\begin{equation}
    SharpeRatio(\mathbf{R}, r_f) = \frac{\overline{\mathbf{R}}-r_f}{\sigma(\mathbf{R})},
\end{equation}

\noindent where $\overline{\mathbf{R}}$ represents the mean of the portfolio returns $\mathbf{R}$, $r_f$ is the risk-free rate, and $\sigma(\mathbf{R})$ is the standard deviation of the portfolio returns $\mathbf{R}$, which measures the volatility of the investment. A higher Sharpe ratio indicates that the investment has a higher return relative to its risk. The Sharpe ratio implementation can be found in Table \ref{code:objectives_and_constraints_part1}.

\subsubsection{Objective: Conditional Value at Risk (CVaR)}
\label{subsubsec:obj_cvar}

The Conditional Value-at-Risk (CVaR) is a risk measure that quantifies the mean of the losses that exceed the Value-at-Risk (VaR) cutoff point:

\begin{equation}
    CVaR(\mathbf{R}, \alpha) = VaR_\alpha + \frac{1}{\alpha}\mathbb{E}[\max(-\mathbf{R}-VaR_\alpha, 0)],\label{eq:cvar}
\end{equation}

\noindent where:

\begin{equation}
    VaR_\alpha = - \inf \{r: P(\mathbf{R} \leq r) \geq \alpha\},
\end{equation}

\noindent where $VaR_\alpha$ is the value-at-risk at the confidence level $\alpha$, which is the minimum value such that the cumulative distribution function (CDF) of the portfolio's returns $\mathbf{R}$, $P(\mathbf{R}\leq r)$, is at least equal to a certain confidence level $\alpha$. The CVaR implementation can be also found in Table \ref{code:objectives_and_constraints_part1}.

\subsubsection{Constraint: Maximum Tracking error allowed}
\label{subsubsec:cons_tracking_error}

The Tracking error (TE) measures the differences between the returns of the portfolio and its benchmark (for example, a reference index) to which it is being compared, as described in the following equation:

\begin{equation}
    TrackingError(\mathbf{R}, \mathbf{I}) = \sigma(\mathbf{R}-\mathbf{I}),
\end{equation}

\noindent where $\mathbf{I}$ represents the index's returns, and $\sigma(\cdot)$ represents the standard deviation. In other words, TE measures the volatility of the difference between the portfolio and index returns. Thus, if we want to limit the maximum TE allowed ($TE_{max}$), we must define the following constraint:

\begin{equation}
    C_{TE} = ReLU(TrackingError(\mathbf{R}, \mathbf{I}) - TE_{max}) \label{eq:cte}
\end{equation}

The Tracking Error implementation in Tensorflow can be found in Table \ref{code:objectives_and_constraints_part1}. This constraint implementation and all the remaining ones can be found in Table \ref{code:objectives_and_constraints_part2}.

\subsubsection{Constraint: UCITS directive}
\label{subsubsec:cons_ucits}

The Undertaking for Collective Investment in Transferable Securities (UCITS) is a set of regulatory rules in the European Union that allows collective investment schemes to operate freely throughout the EU on the basis of a single authorization from one member state. Essentially, UCITS constraints are designed to ensure that funds are not overly concentrated in a small number of assets or sectors. For simplicity, we focus on the following two constraints: (1) The weight of each asset must be below $10\%$, and (2) the sum of the weights exceeding a lower limit ($5\%$) must not exceed an upper limit ($40\%$). This way, we must define two additional constraints:

\begin{equation}
    C_{10\%} = \sum_{i=1}^n ReLU(\mathbf{w}_i - 0.1), \label{eq:ucits1}
\end{equation}

\begin{equation}
    C_{5-40\%} = ReLU\left( (\sum_{i=1}^n{\mathbf{\hat{w}}_i}) - 0.4\right), \label{eq:ucits2}
\end{equation}

\noindent where:

\begin{equation}
    \hat{\mathbf{w}} = \mathbf{w} \odot \hat{\sigma}(\mathbf{w} - 0.05),
\end{equation}

\noindent where $\odot$ represents the element-wise product, and $\hat{\sigma}$ is the $round$ sigmoid function described in equation \ref{eq:round_sigmoid}.

\subsubsection{Constraint: Active weights over a minimum}
\label{subsubsec:cons_weights_min}

The following constraint forces the weights of the active assets to exceed a minimum value $m$. Thus, we must penalize those weights in the range $(0, m)$, so we define a mask that filters the corresponding weights and then apply a penalty, as shown in the following equations:

\begin{equation}
    \mathbf{\hat{w}} = \mathbf{w} \odot \hat{\sigma}(m-\mathbf{w})
\end{equation}

\begin{equation}
    C_{min} = \sum_{i=1}^n \hat{w}_i \label{eq:weight_over_min_constraint}
\end{equation}

\subsubsection{Constraint: Number of active assets within a range}
\label{subsubsec:cons_range}

The following constraint requires the number of active assets to be within the range $(low, high)$. To solve this, we must define a mask $\mathbf{m}$ that allows us to count the number of selected assets, and then define a penalty by combining two thresholds, as shown in the following equations:

\begin{equation}
    \mathbf{m} = \hat{\sigma}(\mathbf{w}),
\end{equation}

\begin{equation}
    C_{range} = ReLU\left((low - \sum_{i=1}^n \mathbf{m}_i)(high - \sum_{i=1}^n \mathbf{m}_i)\right), \label{eq:range_constraint}
\end{equation}

\noindent where $low$ and $high$ define the allowed range, $\mathbf{m}$ defines the vector mask ($\mathbf{m}_i=0$ if $w_i < 0$, else $\mathbf{m}_i=1$).

\subsubsection{Constraint: Sum of the weights of the assets selected by a mask}
\label{subsubsec:cons_assetmask}

The following constraint requires that the total weight assigned to the assets belonging to each predefined group (or mask) does not exceed a specified threshold. Each mask may contain an arbitrary number of assets and multiple masks can be defined. The combined weight across all masks must remain within the feasible budget, typically constrained by the total portfolio weight (e.g., not exceeding 1).

Let $\mathbf{M} \in \{0,1\}^{n \times m}$ be a binary mask matrix, where $n$ is the number of assets and $m$ the number of masks. An element $\mathbf{M}_{ij} = 1$ if the $i$th asset belongs to the $j$th mask, and $0$ otherwise. Also, let $\mathbf{m}_{\text{max}} \in \mathbb{R}^m$ represent the maximum total weight allowed for each mask. The penalty for deviations from the constraint is given by:


\begin{equation}
    C_{\text{mask}} = \sum_{j=1}^{m} | \mathbf{m}_{\text{max},j} - \sum_{i=1}^{n} \mathbf{w}_i \mathbf{M}_{ij} | 
    \label{eq:mask}
\end{equation}

Note that the sum of the maximum total weights allowed for each mask must not exceed 1, although it may be lower than 1 ($\sum_{i=1}^m\mathbf{m}_{max, i} \leq 1$). This allows the optimizer the flexibility to allocate the remaining portion as it deems optimal.

\subsection{Experiments}
\label{subsec:experiments}

For the experiments, we use the assets from the S\&P 500 index, which includes approximately 500 of the largest publicly traded companies in the U.S., offering a diverse set of assets that are ideal for testing and evaluating optimization models. This index is widely recognized as one of the most common benchmarks in portfolio optimization and finance in general, and is used frequently by investors, analysts, and researchers alike. The S\&P 500 provides a sufficient number of assets to build different scenarios under various constraints and objectives; therefore, it is a natural choice for benchmarking MPO problems. Furthermore, it is readily available to anyone through platforms like Yahoo Finance, which offers easy access to historical data through its libraries. In this scenario, for the sake of evaluation of our gradient descent optimizer benchmark, we propose six different situations where the only difference between the optimizers is the loss function.

\subsubsection{Case 1: Maximize Sharpe ratio}
\label{subsubsec:max_sharpe}

For the first case, we consider a single-objective optimization problem closely related to the Markowitz model (see Equation \ref{eq:markowitz}). Specifically, the problem seeks to maximize the Sharpe ratio, as detailed in Section \ref{subsubsec:obj_sharpe}. The loss function that optimizes the Sharpe ratio is defined as:

\[
    L(\mathbf{z}) = -\frac{\overline{\mathbf{R}} - r_f}{\sigma(\mathbf{R})}
\]

\noindent where the objective function $L(\mathbf{z})$ is designed for minimization, so its negative formulation ensures that maximizing the Sharpe ratio is equivalent to minimizing \(L(\mathbf{z})\). This simple scenario allows us to compare the results of our proposal with those achieved by the famous CVXPY convex optimization library \citep{cvxpy1,cvxpy2}.

\subsubsection{Case 2: Minimize CVaR}
\label{subsubsec:min_cvar}

The second scenario is also a single-objective optimization, but this time the objective is to minimize CVaR (see section \ref{subsubsec:obj_cvar}), that means to minimize extreme losses. The loss function is the following:

\begin{equation}
    L(\mathbf{z}) = CVaR(\mathbf{R}, \alpha), 
\end{equation}

\noindent where $\mathbf{R}$ is the series of returns in the portfolio and $\alpha$ is the confidence level to calculate the VaR cutoff point (typically $\alpha=0.05$). In this case, the objective function \(L(\mathbf{z})\) minimizes CVaR, which is equivalent to minimizing extreme losses. Once again, this single-objective scenario will be useful for comparing the benchmark with exact algorithms, implemented in this case by the SKFOLIO library \citep{skfolio}.

\subsubsection{Case 3: Minimize CVaR with UCITS constraints}
\label{subsubsec:min_cvar_ucits}

In this scenario, our objective is to minimize CVaR while simultaneously satisfying a simplified version of the UCITS constraints, as detailed in Section \ref{subsubsec:cons_ucits}. This is our first multi-objective portfolio optimization, where some constraints are used. The loss function is defined as follows:

\begin{equation}
    L(\mathbf{z}, \lambda_{10\%}, \lambda_{5-40\%}) = CVaR(\mathbf{w}, \alpha) + \lambda_{10\%}\, C_{10\%} + \lambda_{5-40\%}\,C_{5-40\%}, \label{eq:loss_ucits}
\end{equation}

\noindent where $\lambda_{10\%}$ and $\lambda_{5-40\%}$ are scaling hyperparameters that quantify the penalty for constraints and the objective. Lower values of these hyperparameters result in more permissive enforcement, while higher values impose stricter penalties. Note that the incorporation of additional constraints is achieved by introducing a new hyperparameter \(\lambda_x\) multiplied by the corresponding constraint \(C_x\).

\subsubsection{Case 4: Minimize CVaR with a tracking error constraint}
\label{subsubsec:min_cvar_te}

In this scenario, our objective is to minimize CVaR while limiting the maximum allowable tracking error (see the details of the constraint in Section \ref{subsubsec:cons_tracking_error}). The loss function is defined as follows:

\begin{equation}
    L(\mathbf{z}, \lambda_{TE}) = CVaR(\mathbf{R}, \alpha) + \lambda_{TE}\, C_{TE},
\end{equation}

\noindent where $\lambda_{TE}$ is the scaling hyperparameter that controls the penalty for exceeding the maximum allowed value $TE_{max}$. If it is critical that the tracking error does not exceed the maximum limit, $\lambda_{TE}$ should be set to a high value. In contrast, if a more flexible approach to this constraint is acceptable, it can be reduced.

\subsubsection{Case 5: Joint optimization of Sharpe ratio and CVaR under Tracking Error, UCITS, Minimum Active Asset Weights, and Active Asset Count Constraints}
\label{subsubsec:max_sharpe_min_cvar_te}

In this case, we pursue two objectives: maximizing the Sharpe ratio and minimizing CVaR, while simultaneously enforcing several constraints. Specifically, the tracking error must not exceed an upper limit, the portfolio must comply with UCITS regulations, each selected asset must meet a minimum weight requirement, and the total number of selected assets must fall within a specified range. Given the dual objectives, they are balanced through lambda hyperparameters, resulting in the following loss function:

\begin{equation}
    L(\mathbf{z}, \lambda_1, \lambda_2, \overline{\lambda}, \overline{C}) = - \lambda_1 \frac{\overline{\mathbf{R}} - r_f}{\sigma(\mathbf{R})} + \lambda_2\, CVaR(\mathbf{R}, \alpha) + \overline{\lambda}\, \overline{C}^T,
\end{equation}

\noindent where $\lambda_1$, $\lambda_2$, and $\overline{\lambda} = \{\lambda_{TE}, \lambda_{10\%}, \lambda_{5-40\%}, \lambda_{min}, \lambda_{range}\}$ balance the importance of objectives and constraints in the loss function, and $\overline{C} = \{C_{TE}, C_{10\%}, C_{5-40\%}, C_{min}, C_{range}\}$ is the vector constraints, whose components are defined in equations \ref{eq:cte}, \ref{eq:ucits1}, \ref{eq:ucits2}, \ref{eq:weight_over_min_constraint}, and \ref{eq:range_constraint}. Note the negative sign preceding the Sharpe ratio term, which reflects its maximization objective. This larger case, compared to the previous cases, serves as an example of how multi-objective optimization can be performed just by adding further objectives or constraints to the loss function.

\subsubsection{Case 6: Minimize Risk with weight constraints in assets subsets}

Finally, we have a single objective: minimize volatility measured as the standard deviation of daily logarithmic returns, while enforcing upper bounds on the total weights assigned to specific subsets of assets (see section \ref{subsubsec:cons_assetmask}). This type of constraint can be particularly useful for limiting the portfolio's exposure to certain sectors or themes.

In this last case, we define four asset masks, each containing ten randomly chosen assets from the investment universe. Each mask represents a subset of assets for which we want to control the aggregate weight. A random maximum weight is assigned to each mask, and the total combined weight allowed in all four masks is 90\% of the portfolio. This allows the optimizer to allocate the remaining 10\% of the portfolio to assets outside of the defined masks. It is important to mention that this constraint does not require the optimizer to assign weights to every asset within each mask, but only enforces that the total weight within each mask does not exceed its respective upper limit. The resulting loss function is defined as:

\begin{equation}
 L(\mathbf{z}, \lambda_{mask}) =\sigma(\mathbf{R}) -\lambda_{mask}\, C_{mask},
\end{equation}

\noindent where $\lambda_{mask}$ balances the importance of the objective and constraint in the loss function, $\mathbf{R}$ denotes the portfolio return, and $C_{mask}$ is the constraint defined in equation \ref{eq:mask}.

Through these six cases, our aim is to demonstrate the flexibility of our optimization framework in handling single- and multi-objective portfolio optimization problems with diverse constraints. The following section shows the effectiveness of our gradient descent optimizer in MPO.

\section{Results and Analysis}
\label{sec:results}

This section presents the experimental evaluation of our framework for MPO problems. We evaluate the performance of the optimizer in the six distinct scenarios detailed in the previous section. The results are analyzed by comparing our gradient descent (GD) optimizer with well-established optimization tools such as CVXPY and SKFOLIO in the simpler scenarios. For more complex constrained cases, we focus on verifying its ability to satisfy regulatory constraints while maintaining strong risk-return profiles. Performance metrics such as Sharpe ratio, Tracking Error, Value-at-Risk (VaR), and Conditional Value-at-Risk (CVaR) are used to evaluate the resulting portfolios. In all cases, the hyperparameters for the optimizer are given. The training dataset is composed by the S\&P 500 Index and its components from 2019-12-31 to 2020-12-31.

\subsection{Case 1: Maximize Sharpe ratio}

\begin{figure}[!bht]
    \centering
    \includegraphics[width=0.7\linewidth]{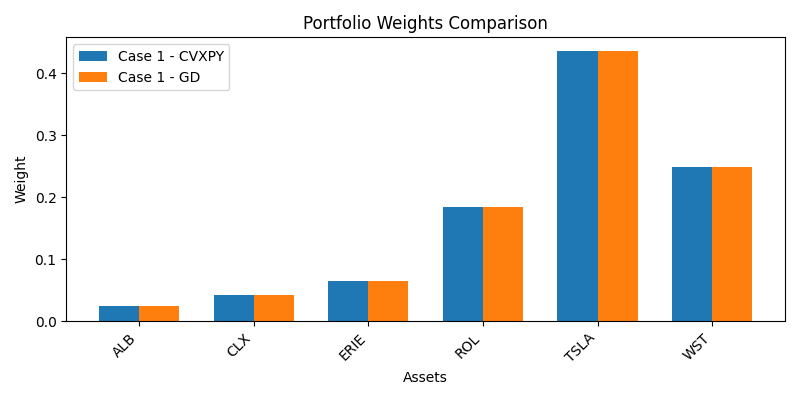}
    \caption{Weights comparison for selected assets (CVXPY vs GD) of S\&P 500, whose weights are greater than $0$.}
    \label{fig:weights_comparison_case1}
\end{figure}

In this first experiment, we compare the exact solution obtained using the CVXPY library \citep{cvxpy1,cvxpy2}, which performs convex optimization, with the results from our gradient descent-based optimizer, trained with a learning rate of $0.01$ for $500$ epochs. Figure~\ref{fig:weights_comparison_case1} presents the portfolio weights resulting from both optimization methods, displaying only the assets with non-zero allocations.

As shown in the figure, the weights for the selected assets are almost equal in both optimizations, with some slight differences shown in Table \ref{tab:weights_comparison_case1}, which shows the exact values.

\begin{table}[!bht]
\centering
\resizebox{\textwidth}{!}{%
\begin{tabular}{lcccccc}
\hline
            & ALB     & CLX     & ERIE    & ROL     & TSLA    & WST     \\
\hline
Case 1 - CVXPY & 0.024792 & 0.041742 & 0.064925 & 0.183647 & 0.436457 & 0.248437 \\
Case 1 - GD    & 0.024794 & 0.041746 & 0.064925 & 0.183647 & 0.436454 & 0.248436 \\
\hline
\end{tabular}%
}
\caption{Weights comparison for selected assets (CVXPY vs GD) of S\&P 500, whose weights are greater than $0$. The weights are almost equal, differing from the 6th decimal.}
\label{tab:weights_comparison_case1}
\end{table}

In this first experiment, we show that our gradient descent optimizer obtains almost the same exact solution as the CVXPY convex optimization. For an additional comparison, we show in Table \ref{tab:metrics_comparison_case1} some performance metrics: the Sharpe ratio of the portfolio, its Tracking Error evaluated over the S\&P 500 index, its value-at-risk, and the conditional value-at-risk, where there are no differences between them.

\begin{table}[!bht]
\centering
\begin{tabular}{lcccc}
\hline
                  & Sharpe   & TrackingError & VaR     & CVaR \\
\hline
Case 1 - CVXPY   & 0.153859 & 0.022139       & 0.044256 & 0.076548      \\
Case 1 - GD      & 0.153859 & 0.022139       & 0.044256 & 0.076548      \\
\hline
\end{tabular}%
\caption{Metrics comparison (CVXPY vs GD) when maximizing Sharpe ratio. The obtained values are totally equal.}
\label{tab:metrics_comparison_case1}
\end{table}

To ensure the robustness of the results, we conducted $K = 100$ random simulations to measure the similarity between both methods. In each simulation, both the set of assets and the training time window are randomly selected, with a universe of 10 assets minimum. For each $k \in \{1, \dots, K\}$, let $\mathbf{u}^{(k)} = (u_1^{(k)}, \dots, u_d^{(k)})$ represent the optimal weight vector obtained from CVXPY, and $\mathbf{v}^{(k)} = (v_1^{(k)}, \dots, v_d^{(k)})$ the corresponding vector from GD. The dimension $d$ denotes the number of assets in the portfolio for simulation $k$.

To measure the discrepancy between both methods, we compute the Euclidean distance between the vectors $u$ and $v$ on average. This metric provides a direct quantification of how close the GD-based and CVXPY-based optimizations are in terms of the overall weight allocation. A low MSE indicates that both methods yield nearly identical portfolio compositions, while a high MSE reveals a significant divergence.

The result of the simulation is an MSE between CVXPY and GD portfolios of $3.6179\times10^{-5}$, which means identical results in practice.

\subsection{Case 2: Minimize CVaR}

\begin{figure}[!bht]
    \centering
    \includegraphics[width=0.7\linewidth]{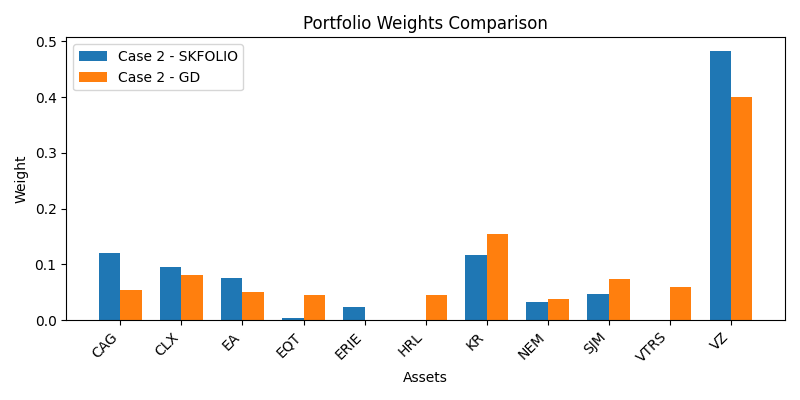}
    \caption{Weights comparison for selected assets (SKFOLIO vs GD).}
    \label{fig:weights_comparison_case2}
\end{figure}

In the second experiment, our objective is to minimize the Conditional Value-at-Risk (CVaR) for the portfolio. We compare exact portfolio allocations and performance metrics obtained by the SKFOLIO optimization method with our gradient descent optimizer, trained with a learning rate of $0.001$ for $2000$ epochs.  Figure~\ref{fig:weights_comparison_case2} shows the portfolio weights for the selected assets as determined by both methods. The exact weight values are provided in Table~\ref{tab:weights_comparison_case2}. 

\begin{table}[!bht]
\centering
\resizebox{\textwidth}{!}{%
\begin{tabular}{lcccccc}
				  & CAG     & CLX     & EA      & EQT     & ERIE    &  HRL   \\
\hline
Case 2 - SKFOLIO  & 0.120858 &  0.095492 &  0.076341 &  0.004152 &  0.024177 &  0.000000 \\
Case 2 - GD       & 0.053394 &  0.081054 &  0.050614 &  0.045719 &  0.000000 &  0.044217 \\
\hline
                  & KR      & NEM     & SJM     & VTRS    & VZ \\
\hline
Case 2 - SKFOLIO  & 0.116439 &  0.032169 &  0.047150 &  0.000000 &  0.483222 \\
Case 2 - GD       & 0.154739 &  0.037236 &  0.073153 &  0.059327 &  0.400548 \\
\hline
\end{tabular}%
}
\caption{Weights comparison for selected assets (SKFOLIO vs GD) of S\&P 500, whose weights are greater than 0 in one option at least.}
\label{tab:weights_comparison_case2}
\end{table}

\begin{figure}[!bht]
    \centering
    \includegraphics[width=0.8\linewidth]{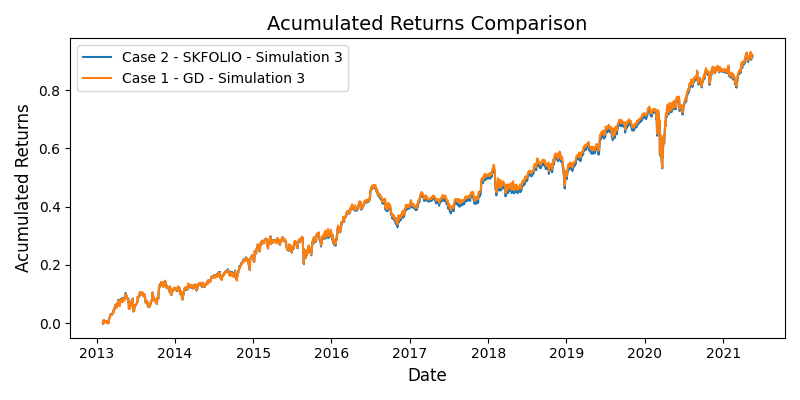}
    \caption{Acummulated returns comparison (SKFOLIO vs GD)}
    \label{fig:MinCVaR_portfolio_acum_returns_comparison}
\end{figure}

As observed in the figure and table, there are clear differences in the individual allocations, including different assets selected and different weights. For example, SKFOLIO chooses Erie Indemnity Company (ERIE) to invest in, while our gradient descent optimizer ignores it and prefers to invest in Hormel Foods Corporation (HRL) and Viatris Inc. (VTRS). In addition, the portfolio weights are different for each optimizer. However, it must also be considered that these 11 assets have been chosen from more than 400 assets in the dataset, and both methods share 8 of them. The key to understanding these differences appears when we show, in addition to the weights, the performance of the portfolios. First, we examine the accumulated returns, as shown in Figure~\ref{fig:MinCVaR_portfolio_acum_returns_comparison}, supporting the similarity in performance between the two portfolios.

Second, we analyze the key performance metrics, shown in Table~\ref{tab:metrics_comparison_case2}, where we include Sharpe ratio, Tracking Error, VaR, and CVaR. The metrics indicate that, while there are differences, the GD optimizer closely approximates the SKFOLIO solution in terms of CVaR, which is the objective of the optimization, showing differences in the 5th decimal.

\begin{table}[!bht]
\centering
\begin{tabular}{lcccccc}
\hline
                  & Sharpe   & TrackingError & VaR     & CVaR \\
\hline
Case 2 - SKFOLIO & 0.006795 & 0.014815 & 0.018191 & 0.027215 \\
Case 2 - GD      & 0.003704 & 0.014573 & 0.018153 & 0.027246 \\
\hline
\end{tabular}%
\caption{Metrics comparison (SKFOLIO vs GD) when minimizing CVaR. The obtained values differ, but the objective to minimize (CVaR) is almost equal, differing in the 5th decimal.}
\label{tab:metrics_comparison_case2}
\end{table}

Although the performance metrics and accumulated returns of both methods appear comparable, especially in terms of CVaR, where the difference is negligible, there is a fundamental reason why the gradient descent optimizer does not exactly replicate the SKFOLIO solution. The CVaR, in its standard form (see section \ref{subsubsec:obj_cvar}), is known to lead to a generally non-convex optimization problem. However, it is possible to transform the problem into a linear program in order to use convex optimization algorithms, like CVXPY or SKFOLIO, and find the global solution. 

Our gradient descent optimizer is an iterative algorithm that solves convex and non-convex problems indistinctly, yet it still achieves results that closely approximate those of the global solution while avoiding the additional complexity in the mathematical transformations. The results shown in table \ref{tab:metrics_comparison_case2} and figure \ref{fig:MinCVaR_portfolio_acum_returns_comparison} are just an example of how our algorithm converges to suboptimal local minima. Considering the following cases, where constrained non-convex multi-objective problems are presented, our gradient descent benchmark is completely valid and reliable.

Consistent with the procedure in Case 1, we ran 100 random simulations and evaluated the mean squared error (MSE) across them. In addition, we have calculated the difference of both the SKFOLIO and GD CVaR of each simulation to evaluate the discrepancy between the exact method (SKFOLIO) and our GD benchmark, since there are small variations in the results shown in figure \ref{fig:weights_comparison_case2}.

The MSE between the weights of the SKFOLIO and GD portfolios was $0.003646$, and the MSE of the CVaR of each of them was $8.7775\times10^{-10}$, which means almost identical weights and the same achievement of the CVaR objective.

\subsection{Case 3: Minimize CVaR with UCITS constraints}

\begin{figure}[!bht]
    \centering
    \includegraphics[width=1\linewidth]{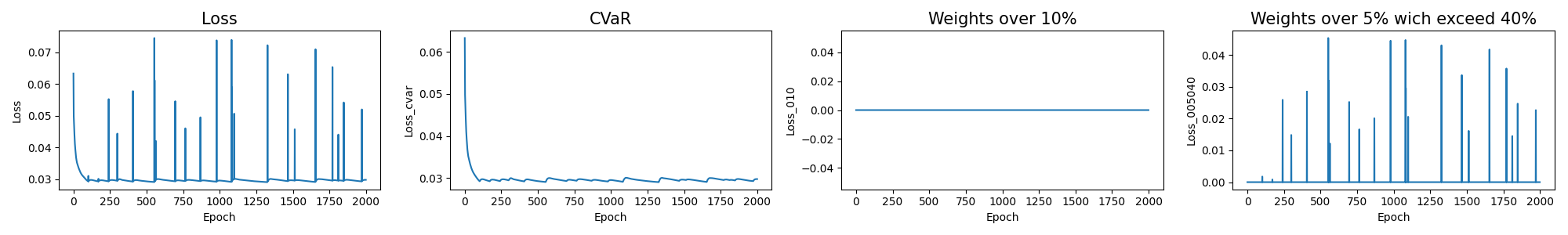}
    \caption{Minimizing CVaR with simplified UCITS constraints training metrics. The first figure (\textit{Loss}) shows the optimization global loss, which follows equation \ref{eq:loss_ucits}, and the remaining show the terms involved in the optimization process.}
    \label{fig:MinCVaR_UCITS_gd_optimizer_training_metrics}
\end{figure}

In this experiment, we extend the CVaR optimization problem by incorporating simplified UCITS constraints, emulating that the portfolio complies with regulatory requirements. In this case, this is a non-DCP (Disciplined Convex Programming) problem, thus CVXPY and SKFOLIO can not find an exact solution, and here is where our benchmark highlights. It was trained with a learning rate of $0.001$ for $2000$ epochs. The objective and constraints hyperparameters were set to $\lambda_{\text{10\%}} = 1.0$, and $\lambda_{\text{5-40\%}} = 1.0$. The training process of the gradient descent-based optimizer is illustrated in Figure~\ref{fig:MinCVaR_UCITS_gd_optimizer_training_metrics}, which shows the evolution of the training metrics and confirms the convergence of the optimization algorithm. From left to right, the first plot (\textit{Loss}) shows the optimization global loss, which follows equation \ref{eq:loss_ucits}, and the remaining show the terms involved in the optimization process. The results shown in the plots represent the product $\lambda_x C_x$.

\begin{figure}[!bht]
    \centering
    \includegraphics[width=1\linewidth]{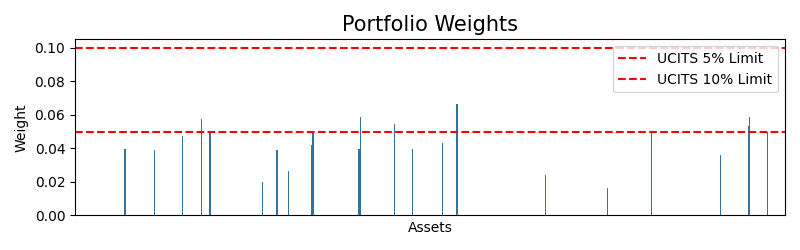}
    \includegraphics[width=1\linewidth]{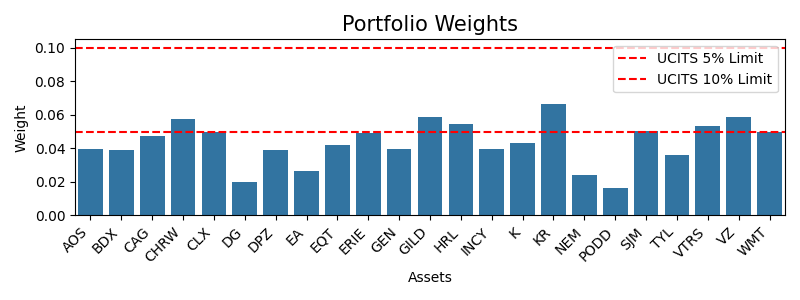}
    \caption{Top figure: Weight assigned to every single asset of the universe along with 5\% and 10\% UCITS limits. Bottom figure: The same portfolio weights for the 23 selected assets. The combined weight of assets exceeding 5\% is 39.98\%.}
    \label{fig:MinCVaR_UCITS_gd_optimizer_portfolio_weights}
\end{figure}

As illustrated in the figures, the optimizer responds when the cumulative weight of assets that exceed 5\% surpasses the 40\% threshold (see the peaks in figure \textit{Weights over 5\% which exceed 40\%}), while the individual weight limit of 10\% is never breached (figure \textit{Weights over 10\%} is always 0). In summary, the result of this optimization is shown in Figure~\ref{fig:MinCVaR_UCITS_gd_optimizer_portfolio_weights}, which shows the portfolio weights along with the 5\% and 10\% UCITS limits. The upper figure shows the weights assigned to every single asset of the more than 400 assets universe, and the lower figure shows the same portfolio weights, but only for the 30 selected assets. Finally, Table~\ref{tab:weights_comparison_case3} presents the exact weights assigned by the GD optimizer to each asset.

\begin{table}[!bht]
\centering
\resizebox{\textwidth}{!}{%
\begin{tabular}{lcccccccc}
& AOS & BDX & CAG & CHRW & CLX & DG & DPZ & EA      \\
\hline
Weight & 0.039855 & 0.038734 & 0.047116 & \textbf{0.057625} & 0.049687 & 0.019987 & 0.039271 & 0.026237  \\
\hline
\\
& EQT & ERIE & GEN & GILD & HRL & INCY & K & KR    \\
\hline
Weight & 0.041741 & 0.049108 & 0.039672 & \textbf{0.058952} & \textbf{0.054397} & 0.039594 & 0.043403 & \textbf{0.066540} \\
\hline
\\
& NEM & PODD & SJM & TYL & VTRS & VZ & WMT \\
\hline
Weight & 0.024005 & 0.016132 & \textbf{0.050218} & 0.035808 & \textbf{0.053264 } & \textbf{0.058796} & 0.049857 \\
\hline
\end{tabular}%
}
\caption{Weights for selected assets. Weights that exceed 5\% are remarked. The sum of remarked weights is 39.9793\%, which is lower than the 40\% threshold of the UCITS constraint.}
\label{tab:weights_comparison_case3}
\end{table}

By analyzing the figures and table, we confirm that the UCITS constraints have been satisfied. The highlighted values in the table correspond to asset allocations exceeding 5\%, but their combined weight does not exceed the limit of 40\% imposed by the UCITS constraint. In addition, we observe that several other assets have allocations close to the 5\% threshold without exceeding it (CAG, CLX, ERIE, WMT), further ensuring compliance with the constraint. For further analysis, we show in Table \ref{tab:metrics_comparison_case2vs3} a comparison of performance when minimizing CVaR with or without the UCITS constraint. 

\begin{table}[!bht]
\centering
\begin{tabular}{lcccccc}
\hline
                  & Sharpe   & TrackingError & VaR     & CVaR \\
\hline
Case 2 - GD      & 0.003704 & 0.014573 & 0.018153 & 0.027246 \\
Case 3 - GD      & 0.019866 & 0.013078 & 0.020722 & 0.029034 \\
\hline
\end{tabular}%
\caption{Metrics comparison when minimizing CVaR. Case 2 does not apply any constraint, while case 3 apply UCITS constraints (see section \ref{subsubsec:cons_ucits}). In both cases, the optimization objective is to minimize CVaR.}
\label{tab:metrics_comparison_case2vs3}
\end{table}

Despite the UCITS constraints in case 3, the resulting CVaR remains very close to that of case 2, demonstrating that the gradient descent optimizer is capable of finding a solution that balances regulatory compliance constraints with the risk minimization objective.

To evaluate the impact of the regularization parameters $\lambda_{\text{10\%}}$ and $\lambda_{\text{5-40\%}}$ on both CVaR and the regulatory compliance of the optimized portfolio, we performed a grid search across a range of values for these hyperparameters. Specifically, $\lambda_{\text{10\%}}$ and $\lambda_{\text{5-40\%}}$ varied independently in the set $\{0.0, 0.001, 0.01, 0.1, 1.0, 10.0\}$.

\begin{figure}[!bht]
    \centering
    \includegraphics[width=0.70\linewidth]{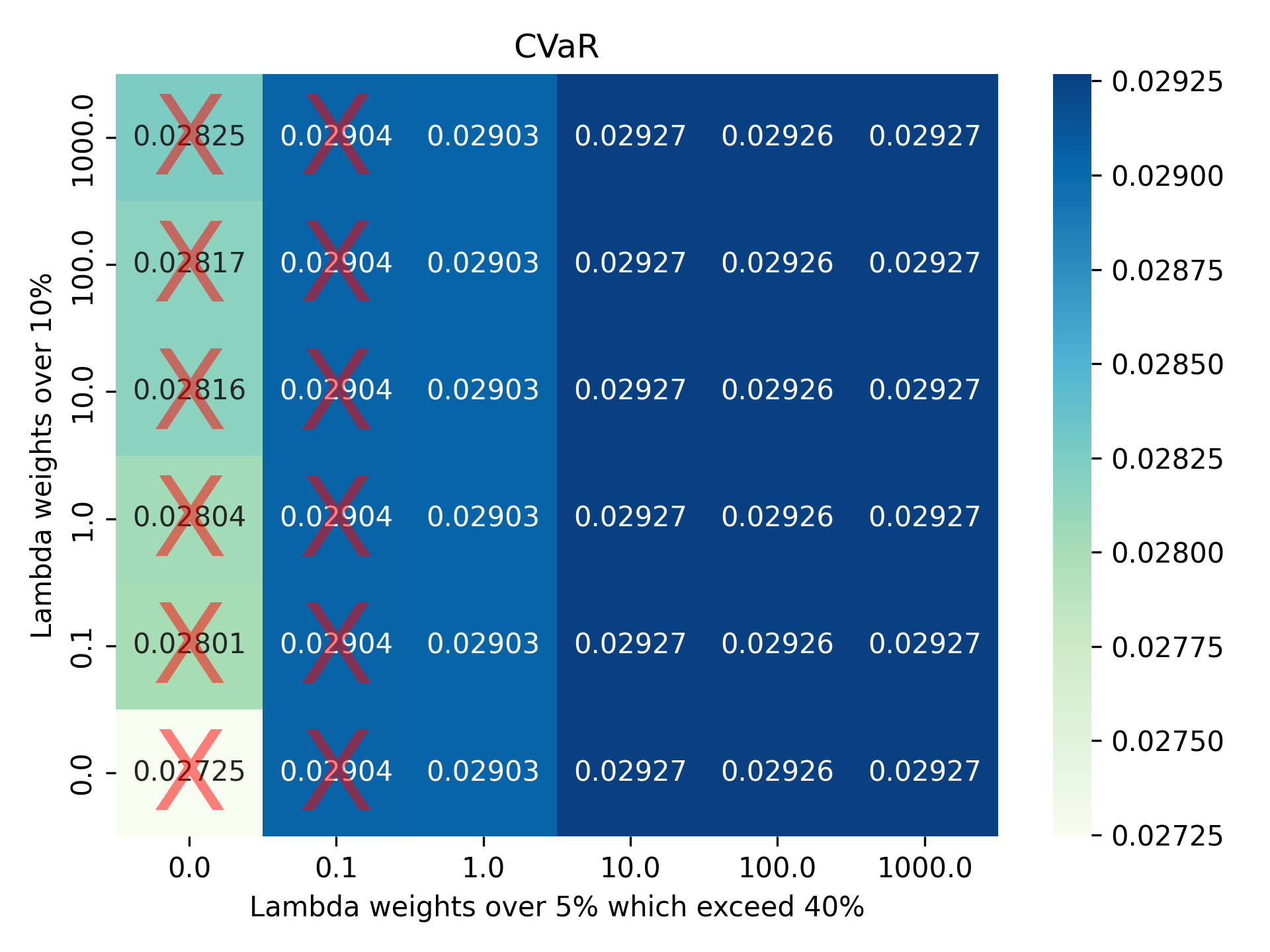}
    \caption{CVaR corresponding to each combination of regularization parameters $\lambda_{\text{10\%}}$ and $\lambda_{\text{5-40\%}}$ during the grid search algorithm. Cells marked with an red X mean that basic UCITS constraints are not satisfied, so that scenario could not be taken into account.}
    \label{fig:MinCVaR_UCITS_gd_optimizer_theoretical_loss}
\end{figure}

Figure \ref{fig:MinCVaR_UCITS_gd_optimizer_theoretical_loss} displays the CVaR of the resulting optimized portfolio for each combination of parameters. Each cell in the heat map represents the CVaR for a specific pair $(\lambda_{\text{10\%}}, \lambda_{\text{5-40\%}})$. Red crosses indicate portfolios that do not comply with basic UCITS constraints and are therefore considered infeasible. Note that the minimum CVaR is obtained when both regularization parameters are set to zero (0.02725). However, this configuration does not satisfy the regulatory requirements. Compliance is achieved only when $\lambda_{\text{5-40\%}}$ takes a value within the range between 1.0 and 10.0. Thus, the optimizer is forced to find the proper tradeoff between minimizing CVaR and satisfying the constraints, with a difference of 0.00178.

\subsection{Case 4: Minimize CVaR with Tracking Error constraint}

\begin{figure}[!bht]
    \centering
    \includegraphics[width=1.33\linewidth]{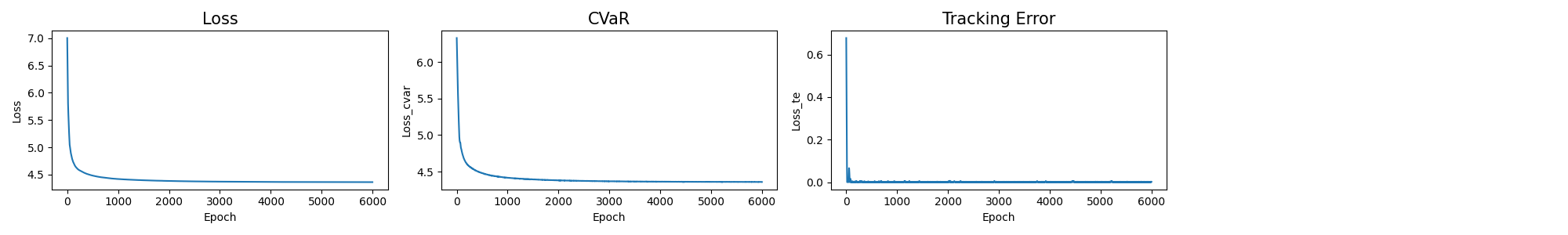}
    \caption{Minimize CVaR with Tracking Error constraint training metrics}
    \label{fig:MinCVaR_Tracking_gd_optimizer_training_metrics}
\end{figure}

Now, the objective is to minimize CVaR while enforcing a maximum Tracking Error constraint. The optimizer was trained with a learning rate of $0.00005$ for $6000$ epochs. The maximum allowed tracking error was set to $0.004$. The objective and constraints hyperparameters were $\lambda_{\text{TE}} = 3$. The training process is illustrated in Figure~\ref{fig:MinCVaR_Tracking_gd_optimizer_training_metrics}, where the evolution of the training metrics indicates a stable convergence of the optimization algorithm under the tracking error constraint.

\begin{figure}[!bht]
    \centering
    \includegraphics[width=0.8\linewidth]{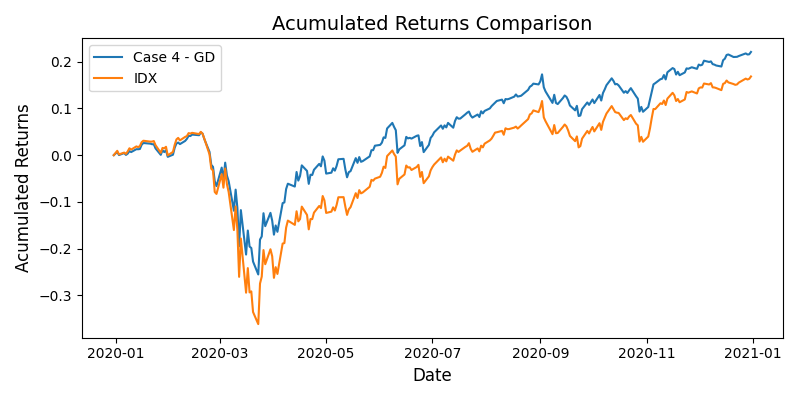}
    \caption{Accumulated returns comparison between the resulting portfolio and the reference index S\&P 500 when minimizing CVaR with Tracking Error constraint of $TE_{max} = 0.004$.}
\label{fig:MinCVaR_Tracking_portfolio_acum_returns_comparison}
\end{figure}

As shown in the figures, both the CVaR and the Tracking Error constraints are effectively satisfied, resulting in a portfolio that not only minimizes tail risk, but also closely follows the benchmark index (S\&P 500) due to an explicit limitation on tracking error equal to 0.4\%, which ensures that the portfolio does not deviate from the reference index. The comparison of accumulated returns in Figure \ref{fig:MinCVaR_Tracking_portfolio_acum_returns_comparison} and Table \ref{tab:metrics_comparison_case4} show this idea, where the resulting portfolio follows the reference index. 

\begin{table}[!bht]
\centering
\begin{tabular}{lcccccc}
\hline
                  & Sharpe   & TrackingError & VaR     & CVaR \\
\hline
S\&P 500 index & 0.016771 & --      & 0.033728 & 0.056878  \\
Case 4 - GD      & 0.031009 & 0.004001      & 0.027631 & 0.043603  \\
\hline
\end{tabular}%
\caption{Metrics comparison (GD vs S\&P 500 index) when minimizing CVaR with a maximum Tracking Error constraint of $TE_{max} = 0.004$. Note that the optimizer reaches an exceeding of $10^{-6}$ with respect to the $TE_{max}$.}
\label{tab:metrics_comparison_case4}
\end{table}

From the figure and the table, we can extract two main ideas. First, by minimizing CVaR and restricting the Tracking Error relative to the index, the portfolio effectively follows the benchmark, while it reduces the exposure to extreme losses. As shown in Table \ref{tab:metrics_comparison_case4}, the gradient descent optimizer portfolio achieves a higher Sharpe ratio, indicating a better risk-return trade-off, while simultaneously lowering CVaR (better performance and lower losses in Figure \ref{fig:MinCVaR_Tracking_portfolio_acum_returns_comparison}). Second, the results in this case show that incorporating a Tracking Error constraint along with the CVaR minimization objective leads to portfolios that maintain competitive risk and return characteristics while ensuring lower extreme losses. This finding underscores the potential of the GD optimizer to achieve a robust balance between constraints and the objective.

\subsection{Case 5: Maximize Sharpe ratio, minimize CVaR with Tracking Error, UCITS, minimum asset weight constraints, and assets number within a certain range}

\begin{figure}[!bht]
    \centering
    \includegraphics[width=1\linewidth]{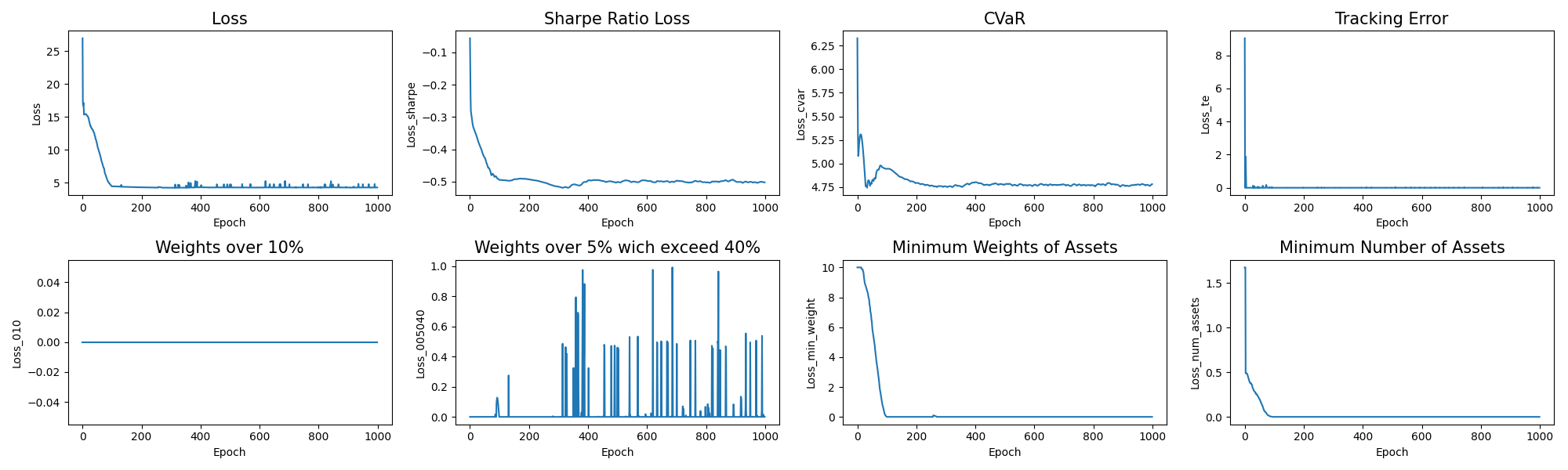}
    \caption{Maximize Sharpe ratio, minimize CVaR with Tracking Error, UCITS, minimum asset weight constraints, and assets number within a range training metrics.}
    \label{fig:MaxSharpe_MinCVaR_Tracking_MinWeight_AssetRange_UCITS_gd_optimizer_training_metrics}
\end{figure}

In this experiment, the goal is to simultaneously maximize the Sharpe ratio and minimize CVaR, while incorporating several practical constraints: a Tracking Error limit, UCITS constraints, a minimum asset weight threshold, and a restriction on the number of selected assets (between 20 and 30). The optimizer was trained with a learning rate of $0.001$ for $1000$ epochs. The optimization hyperparameters were set to $\lambda_{\text{CVaR}} = 100$, $\lambda_{\text{Sharpe}} = 10$, $\lambda_{\text{TE}} = 0.004$, $\lambda_{\text{10\%}} = 10$, $\lambda_{\text{5-40\%}} = 10$, $\lambda_{min} = 10$, and $\lambda_{range} = 0.00001$. The maximum Tracking Error is set to $0.004$ and the minimum weight an asset could have is $1\%$. Figure~\ref{fig:MaxSharpe_MinCVaR_Tracking_MinWeight_AssetRange_UCITS_gd_optimizer_training_metrics} shows the training process, where the global loss is minimized while balancing Sharpe ratio and CVaR.

The figure shows the convergence behavior of the gradient descent optimizer under the complex multi-objective setup. During the epochs, it can be observed that the overall loss decreases steadily, indicating that the optimizer is successfully balancing the objectives (maximizing the Sharpe ratio and minimizing CVaR) while satisfying the additional constraints (tracking error, UCITS, minimum asset weight, and asset count). 

\begin{figure}[!bht]
    \centering
    \includegraphics[width=1\linewidth]{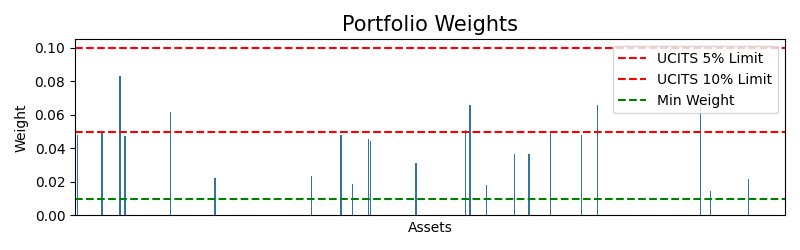}
    \includegraphics[width=1\linewidth]{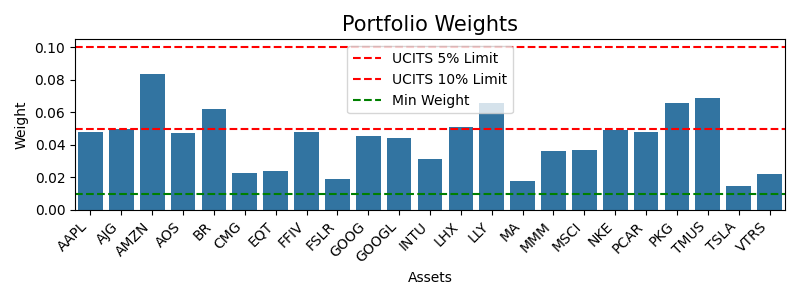}
    \caption{Top figure: Selected assets from the whole S\&P 500 universe. Bottom figure: Portfolio weights for the 23 selected assets. The minimum weight threshold is at 1\%. The allowed range of number of selected assets is between 20 and 30. The sum of weights above 5\% is 39.66\%}
    \label{fig:MaxSharpe_MinCVaR_Tracking_MinWeight_AssetRange_UCITS_gd_optimizer_portfolio_weights}
\end{figure}

In addition Figure~\ref{fig:MaxSharpe_MinCVaR_Tracking_MinWeight_AssetRange_UCITS_gd_optimizer_portfolio_weights} provides a dual perspective. In the top figure, the weights allocated across the entire S\&P 500 universe are shown, highlighting the asset selection process. In the bottom figure, we zoom in on the 23 selected assets. Here, each asset receives at least a 1\% allocation (meeting the minimum weight constraint, represented in green), and the cumulative weight of assets with allocations above 5\% totals 39.66\%. This layout confirms that the UCITS constraints are respected and that the portfolio is diversified and focused.

\begin{figure}[!bht]
    \centering
    \includegraphics[width=0.8\linewidth]{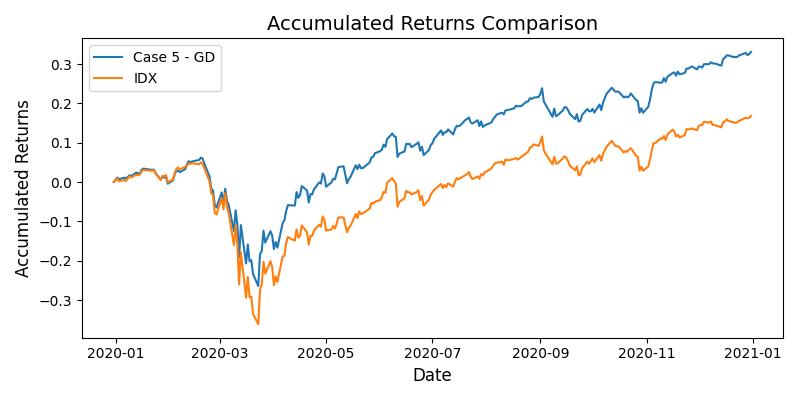}
    \caption{Accumulated returns comparison between portfolio optimized using Gradient Descent and the reference index S\&P 500 in 2020 when facing case 5: maximize Sharpe ratio, minimize CVaR, and adding the Tracking Error, UCITS, minimum asset weight, and assets number within a certain range constraints.}
    \label{fig:MaxSharpe_MinCVaR_Tracking_MinWeight_AssetRange_UCITS_gd_optimizer_portfolio_acum_returns}
\end{figure}

Finally, Figure~\ref{fig:MaxSharpe_MinCVaR_Tracking_MinWeight_AssetRange_UCITS_gd_optimizer_portfolio_acum_returns} compares the performance of the portfolio against the S\&P 500 index. The GD portfolio closely tracks the benchmark while exhibiting a superior risk-return profile, as indicated by the improved Sharpe ratio and lower CVaR. This is a direct consequence of the Tracking Error constraint and the optimization objectives (maximizing the Sharpe ratio and minimization of CVaR). Table~\ref{tab:metrics_comparison_case5} shows the performance metrics of the portfolio compared to S\&P 500.

\begin{table}[!bht]
\centering
\begin{tabular}{lcccccc}
\hline
                       & Sharpe   & TrackingError & VaR      & CVaR   \\
\hline
S\&P 500 Index & 0.016771 & --            & 0.033728 & 0.056878  \\
Case 4 - GD             & 0.031009 & 0.004001      & 0.027631 & 0.043603  \\
Case 5 - GD            & 0.051612 & 0.003996       & 0.031112 & 0.047519  \\
\hline
\end{tabular}%
\caption{Comparison of performance metrics for Cases 4, 5 and S\&P 500 Index.}
\label{tab:metrics_comparison_case5}
\end{table}

Compared to Case 4, where the objective was solely to minimize CVaR under a maximum Tracking Error constraint, Case 5 also introduces the maximization of the Sharpe ratio, while also satisfying other constraints. This approach leads to a significant improvement in performance. As shown in the table, the portfolio in Case 5 achieves a higher Sharpe ratio (0.051612 vs 0.031009) in balance with a slight loss with CVaR (0.047519 vs 0.043603), this is a reasonable trade-off considering the notable increase in risk-adjusted return. Moreover, both cases accomplish with the Tracking Error constraint, confirming that the performance gain in Case 5 is not a result of increased deviation from the benchmark. We refer to the figures \ref{fig:MinCVaR_Tracking_portfolio_acum_returns_comparison} and \ref{fig:MaxSharpe_MinCVaR_Tracking_MinWeight_AssetRange_UCITS_gd_optimizer_portfolio_acum_returns} in order to visually compare the performance of both portfolios.

Overall, the figures and performance metrics together illustrate that the multi-objective and multi-constraint optimization framework effectively constructs a portfolio with balanced risk and return, ensuring compliance with regulations. The lambda hyperparameters balance the trade-off between different objectives and constraints, being the final user who decides which values work better for his investment strategy.

\subsection{Case 6: Minimize Risk with weight constraints in assets subsets}

In this last case, the goal is to minimize portfolio risk, measured by the portfolio’s standard deviation, while enforcing specific weight constraints on four predefined asset subsets (masks). The optimizer was trained with a learning rate of $0.0009$ for $8000$ epochs. The hyperparameters were set to $\lambda_{mask} = 0.1$. The masks were created by randomly choosing 10 assets per group without replacement, and assigning maximum allowed weights to each. The selected assets per mask are listed in Table~\ref{tab:asset_indexes_case6}, and the masks weights constraints are shown in Table~\ref{tab:mask_weight_constraints_case6}. Note that the sum of weights of all masks is 90\%, giving the optimizer flexibility to allocate the remaining 10\% outside the defined masks.

\begin{table}[!bht]
\centering
\small
\begin{tabular}{lcccccccccc}
\hline
 & 1 & 2 & 3 & 4 & 5 & 6 & 7 & 8 & 9 & 10 \\
\hline
Mask 1 & RVTY & AVY  & TJX  & CL   & AMZN & TSCO & UDR  & AWK  & HIG  & MTB  \\
Mask 2 & KMB  & ATO  & HAL  & MO   & GE   & BK   & CPRT & MRK  & INTC & RTX  \\
Mask 3 & NSC  & IPG  & GILD & SYY  & OXY  & SO   & O    & IT   & ROP  & CHD  \\
Mask 4 & PPG  & EXPE & DLR  & PNC  & LH   & MDLZ & PCG  & FSLR & HBAN & ARE  \\
\hline
\end{tabular}%
\caption{Asset indexes randomly chosen without replace for each mask.}
\label{tab:asset_indexes_case6}
\end{table}

\begin{table}[!bht]
\centering
\begin{tabular}{lcccc}
\hline
                       & Mask 1 & Mask 2 & Mask 3 & Mask 4 \\
\hline
Weight & 0.27816742 & 0.40033937 & 0.17409502 & 0.04739819 \\
\hline
\end{tabular}%
\caption{Mask weight constraints randomly selected. The sum of the wights of all masks is 0.9.}
\label{tab:mask_weight_constraints_case6}
\end{table}

\begin{figure}[!bht]
    \centering
    \includegraphics[width=1.33\linewidth]{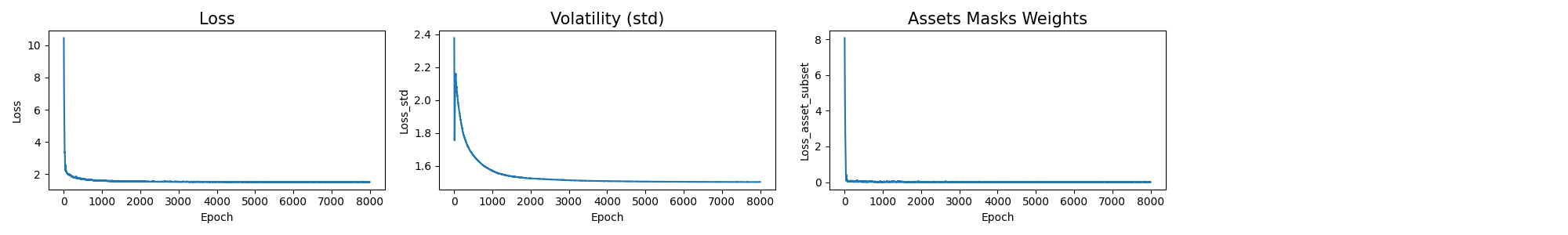}
    \caption{Minimize Risk with weight constraints in assets subsets training metrics}
    \label{fig:MinStd_AssetSubsetWeight_gd_optimizer_training_metrics}
\end{figure}

Figure~\ref{fig:MinStd_AssetSubsetWeight_gd_optimizer_training_metrics} shows the training metrics. The loss, volatility (standard deviation), and mask-related penalty terms all converge smoothly, suggesting that the optimizer effectively balances the goal of risk minimization with the satisfaction of the mask constraints.

\begin{figure}[!bht]
    \centering
    \includegraphics[width=0.49\linewidth]{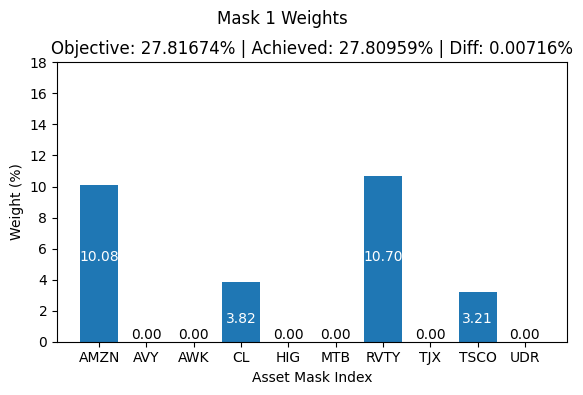}
    \includegraphics[width=0.49\linewidth]{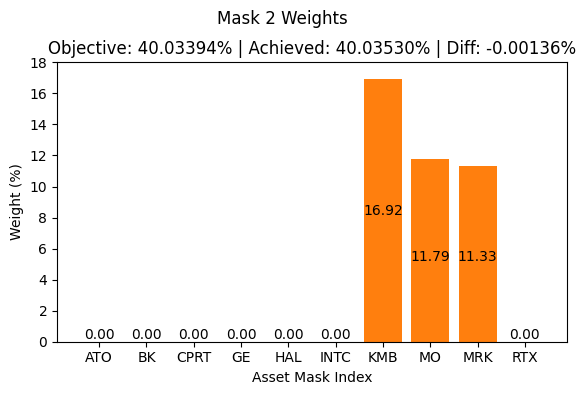}
    \includegraphics[width=0.49\linewidth]{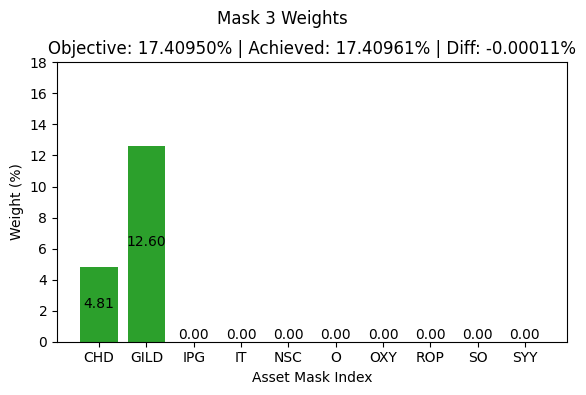}
    \includegraphics[width=0.49\linewidth]{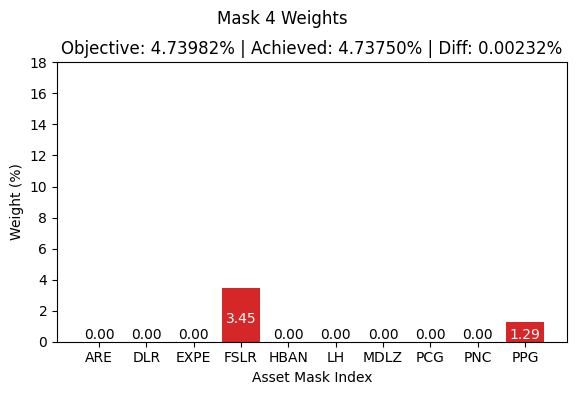}
    \caption{Asset weights selected by each Mask, showing the target objective for the optimizer and the value achieved.}
    \label{fig:case6_mask_1_weights}
\end{figure}

Figure~\ref{fig:case6_mask_1_weights} illustrates how each of the four asset masks distributes its weights. Each subplot includes the maximum allowed weight (objective), the actual accumulated weight achieved by the optimizer, and the difference. This shows that the model adheres tightly to the constraints, and that all masks achieve their targets with negligible deviations (below 0.01\%). Note that the optimizer does not assign weight to every asset included in each mask.

\begin{figure}[!bht]
    \centering
    \includegraphics[width=1\linewidth]{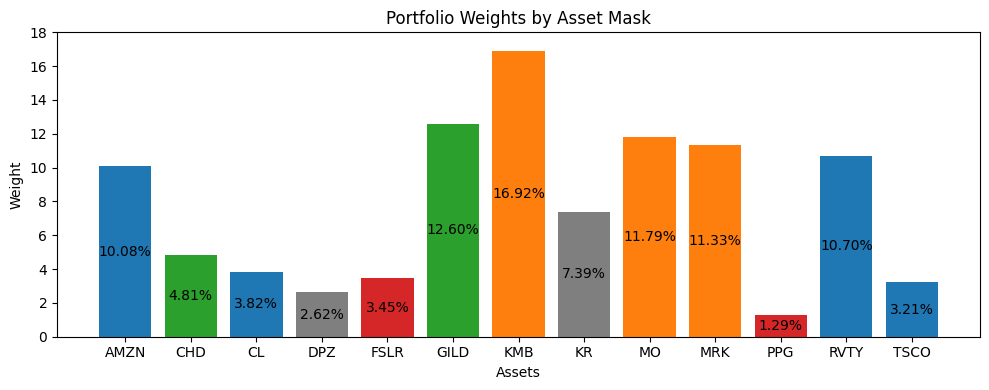}
    \caption{Portfolio weights for 13 selected assets.}    \label{fig:MinStd_AssetSubsetWeight_gd_optimizer_portfolio_weights_no_zeros}
\end{figure}

Finally, Figure~\ref{fig:MinStd_AssetSubsetWeight_gd_optimizer_portfolio_weights_no_zeros} displays the optimized portfolio weights, highlighting the 13 assets with non-zero allocations. While the selected assets are drawn from different masks, not every asset within each mask is used. Notably, the optimizer has assigned weights of 2.61\% to DPZ and 7.39\% to KR, together accounting for exactly 10\% of the portfolio—matching the portion of total weight not constrained by the predefined masks.

Overall, this case confirms that the proposed optimization setup can effectively reduce portfolio volatility while adhering to strict, mask-based allocation constraints. This kind of setup is especially useful in scenarios where asset exposures must follow sector, thematic, or regulatory grouping structures.

\section{Conclusion and Future Work}
\label{sec:conclusion}

In this work, we have presented a framework for multi-objective portfolio optimization (MPO) based on gradient descent with automatic differentiation in Tensorflow. Rather than proposing a novel optimization strategy or aiming to outperform existing portfolio construction techniques, our goal has been to present a flexible, scalable, and extensible tool that can support a wide range of financial objectives and constraints. This benchmark will allow both researchers and practitioners to model complex investment scenarios in a straightforward and customizable manner.

Our experimental evaluation across six scenarios, ranging from single-objective optimization to highly constrained multi-objective cases, shows that the proposed framework performs consistently well. With this optimizer, one can explore the effect of the relative importance of each objective and constraint in multi-objective strategies. This capability of customizing the optimizer is completely useful in real-world financial applications, and the results show that the multi-objective and multi-constraint optimization framework effectively build portfolios with balanced risk and return, ensuring the constraints. The lambda hyperparameters in the loss function balance the trade-off between different objectives and constraints, where the final user decides which values work better for his investment strategy.

This work opens up several promising directions for further research. First, a natural extension would be to perform a comparison with other metaheuristic methods, such as genetic and swarm algorithms. This evaluation would deepen the strengths of gradient-based optimization in terms of convergence speed, performance, and solution quality. Second, our gradient-based approach allows the optimization of multiple scenarios at the same time. Investigating whether training in batches across multiple temporal windows would help to discover globally suboptimal portfolios that could perform robustly in all scenarios at once. In addition, exploring the incorporation of forward-looking models, such as Black-Litterman, or integrating risk-based frameworks like Risk Parity, would guide better allocation decisions. With this work, we hope this benchmark will serve as a foundation for research in portfolio optimization.

\section*{Acknowledgements}

This research was supported by grant PID2023-149669NB-I00 (MCIN/AEI and ERDF - ``A way of making Europe'').

\bibliographystyle{elsarticle-harv} 
\bibliography{mybib}


\newpage

\appendix

\section{Softmax vs Sparsemax}
\label{ap:softmax_vs_sparsemax}

In this paper, we explore the use of the Softmax and Sparsemax functions as mechanisms to assign portfolio weights in the context of optimization algorithms. These functions are evaluated based on their ability to enforce non-negativity and normalization constraints while enabling sparsity in the solution.

The \textbf{Softmax function} transforms an input vector $\textbf{z} \in \mathbb{R}^K$ into a probability distribution:

$$
\sigma(\textbf{z})_j = \frac{e^{z_j}}{\sum_{k=1}^K e^{z_k}} \quad \text{for } j = 1, \ldots, K
$$

To prevent numerical overflow, the function is often stabilized using a constant $M = \max(\textbf{z})$:

$$
\sigma(\textbf{z})_j = \frac{e^{z_j - M}}{\sum_{k=1}^K e^{z_k - M}}
$$

This ensures that the output vector has all elements strictly positive and sums to one. However, a key drawback in financial applications is that Softmax tends to allocate non-zero weights to all elements, which is not ideal when sparsity is desired in asset allocation.

\textbf{Sparsemax Function}: As a remedy to this limitation, the Sparsemax function, introduced by \cite{sparsemax_martins16}, offers a projection onto the probability simplex that can yield exact zeros.

The resulting vector maintains the key properties of non-negativity and summing to one, but unlike Softmax, allows for true sparsity, assigning exactly zero weights to less relevant elements.

\begin{figure}[!bht]
    \centering
    \includegraphics[width=1\linewidth]{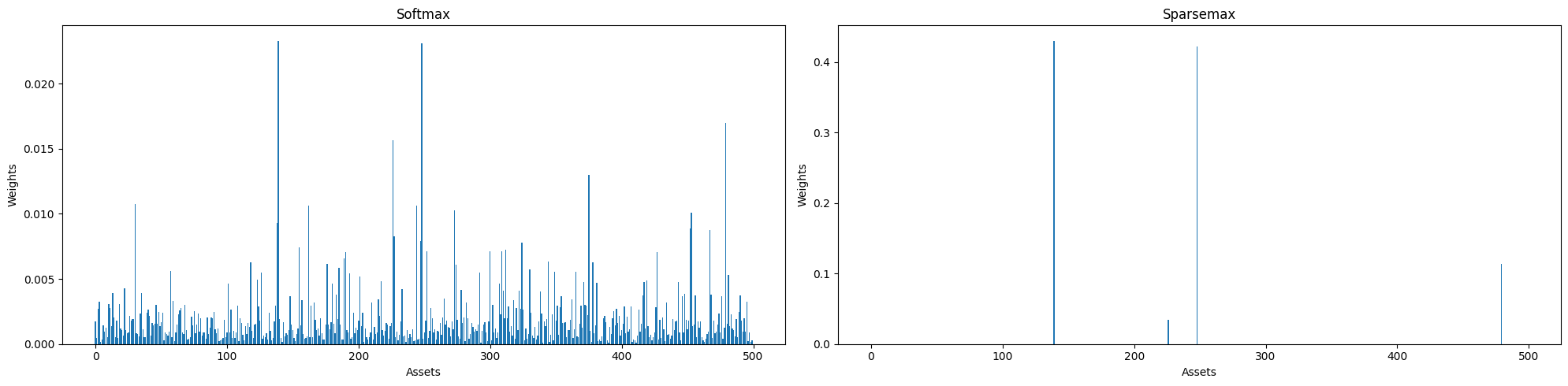}
    \caption{Softmax vs Sparsemax.}
    \label{fig:Softmax_vs_Sparsemax}
\end{figure}

Figure \ref{fig:Softmax_vs_Sparsemax} shows that Sparsemax can output weight vectors with many exact zeros, promoting interpretable and concentrated portfolios.  When dealing with hundreds of assets, Sparsemax avoids capital dilution by zeroing out low-scoring assets. Unlike ad-hoc truncation and renormalization applied to Softmax outputs, Sparsemax respects the original structure of the optimized vector.

In the context of portfolio optimization, the Softmax function is a convenient but dense mapping. Sparsemax offers an alternative that encourages sparsity, improving both interpretability and practicality in large-scale asset allocation.

\newpage

\section{Differentiation techniques in Tensorflow}
\label{ap:tf_differentiation_techniques}

\begin{table}[!bht]
    \centering
    \begin{lstlisting}[label=lst:python_example]

        def round_sigmoid(decimals=0):
            @tf.custom_gradient
            def rs(x):
                z = tf.nn.sigmoid(x)
                scale = 10**decimals
                output = tf.math.round(z * scale) / scale
            
                def backward(dy):
                    custom_grad = z * (1 - z)
                    return dy * custom_grad
                
                return output, backward
            return rs

    \end{lstlisting}
    \caption{Tensorflow implementation of the $round$ of the sigmoid function, following the equation \ref{eq:round_sigmoid} and its derivative (eq. \ref{eq:derivative_round_sigmoid}).}
    \label{code:round_sigmoid}
\end{table}

\begin{table}[!bht]
    \centering
    \begin{lstlisting}[label=lst:python_example]

        def mask_lower_than(x, threshold, epsilon=1e-16):
            return round_sigmoid()(x - threshold - epsilon)

        def mask_greater_than(x, threshold, epsilon=1e-16):
            return round_sigmoid()(threshold + epsilon - x)

    \end{lstlisting}
    \caption{Tensorflow implementation of the mask described in equation \ref{eq:round_sigmoid} using the round sigmoid function (\textit{mask\_lower\_than}) and its analogous \textit{mask\_greater\_than}.}
    \label{code:mask_threshold}
\end{table}

\newpage

\section{Constraints implementation in Tensorflow}
\label{ap:tf_constraints}

\begin{table}[!bht]
    \centering
    \begin{lstlisting}[label=lst:python_example]

        def SharpeRatio(x, r_f=0.0):
            return (tf.math.reduce_mean(x) - r_f) / tf.math.reduce_std(x)

        def VaR(x, alpha = 0.05):
            return -tfp.stats.percentile(x, q=alpha * 100)

        def CVaR(x, alpha = 0.05):
            var = VaR(x, alpha=alpha)
            shortfall = tf.nn.relu(-x - var_alpha)
            mean_shortfall = tf.math.reduce_mean(shortfall) / alpha
            return var + mean_shortfall

        def TrackingError(x, y):
            return tfp.stats.stddev(x - y)

        def Std(x):
            return tf.math.reduce_std(x)

    \end{lstlisting}
    \caption{Tensorflow implementation of portfolio risk and performance objectives. \textit{SharpeRatio} maximizes return-to-risk efficiency, \textit{VaR} and \textit{CVaR} estimate downside risk at a given confidence level, \textit{TrackingError} measures the deviation from a reference benchmark, and \textit{Std} measures the volatility of the portfolio returns.}
\label{code:objectives_and_constraints_part1}
\end{table}

\begin{table}[!bht]
    \centering
    \begin{lstlisting}[label=lst:python_example]

        def ConstraintUCITS_1(w):
            return tf.math.reduce_sum(exceeding_threshold(w, 0.1))

        def ConstraintUCITS_2(w):
            mask = mask_lower_than(w, 0.05)
            return exceeding_threshold(tf.math.reduce_sum(w * mask) - 0.4)

        def ConstraintTrackingError(x, y, TE_max):
            return exceeding_threshold(TrackingError(x, y), TE_max)

        def ConstraintMinWeights(w, min_value):
            mask = mask_greater_than(w, min_value)
            return tf.math.reduce_sum(w * mask)

        def ConstraintRange(w, low, high):
            mask = mask_lower_than(w, 0.0)
            lower = low - tf.math.reduce_sum(mask)
            higher = high - tf.math.reduce_sum(mask)
            return exceeding_threshold(lower * higher)

        def ConstraintSubsets(w, M, m):
            return tf.math.reduce_sum(tf.abs((m - w @ M)))

    \end{lstlisting}
    \caption{Tensorflow implementation of portfolio constraints. These include simplified UCITS regulations (\textit{ConstraintUCITS\_1} and \textit{ConstraintUCITS\_2}), Tracking Error cap (\textit{ConstraintTrackingError}), minimum asset weights (\textit{ConstraintMinWeights}), number of assets within a valid range (\textit{ConstraintRange}), and mask-based subset constraints (\textit{ConstraintSubsets}).}
    \label{code:objectives_and_constraints_part2}
\end{table}

\end{document}